\documentclass[aps,prb,twocolumn,groupedaddress,amsmath,showpacs]{revtex4}

\usepackage{graphicx}
\usepackage{bm}
\usepackage{here}

\newcommand{\bra}[1]{\langle #1|}
\newcommand{\ket}[1]{|#1\rangle}
\newcommand{\braket}[2]{\langle #1|#2\rangle}

\newcommand{\rhombus}{% rhombus for triangular QDM
  \put(0,0){\circle*{4}}
  \put(20,0){\circle*{4}}
  \put(10,17.3){\circle*{4}}
  \put(30,17.3){\circle*{4}}
}

\newcommand{\rhombi}{\unitlength0.04em % QDM-triangular-zero
 \begin{minipage}{35\unitlength}
 \begin{center}
 \begin{picture}(30,17)
  \rhombus
  \put(0,0){\line(1,0){20}}
  \put(10,17.3){\line(1,0){20}}
  \qbezier(0,0)(5,8.65)(10,17.3)
  \qbezier(20,0)(25,8.65)(30,17.3)
 \end{picture}
 \end{center}
 \end{minipage}
}

\newcommand{\QDMtz}{\unitlength0.04em % QDM-triangular-zero
 \begin{minipage}{35\unitlength}
 \begin{center}
 \begin{picture}(30,17)
  \rhombus
  \put(0,0){\line(1,0){20}}
  \put(10,17.3){\line(1,0){20}}
 \end{picture}
 \end{center}
 \end{minipage}
}

\newcommand{\QDMts}{\unitlength0.04em % QDM-triangular-sixty
 \begin{minipage}{35\unitlength}
 \begin{center}
 \begin{picture}(30,17)
  \rhombus
  \qbezier(0,0)(5,8.65)(10,17.3)
  \qbezier(20,0)(25,8.65)(30,17.3)
 \end{picture}
 \end{center}
 \end{minipage}
}

\newcommand{\triup}{
    \put(0,0){\circle*{4}}
    \put(20,0){\circle*{4}}
    \put(10,17.3){\circle*{4}}
}

\newcommand{\tridown}{
    \put(20,0){\circle*{4}}
    \put(10,17.3){\circle*{4}}
    \put(30,17.3){\circle*{4}}
}

\newcommand{\QDMtupa}{\unitlength0.04em
 \begin{minipage}{35\unitlength}
 \begin{center}
 \begin{picture}(30,17)
    \triup
    \put(0,0){\line(1,0){20}}
 \end{picture}
 \end{center}
 \end{minipage}
}

\newcommand{\QDMtupb}{\unitlength0.04em
 \begin{minipage}{35\unitlength}
 \begin{center}
 \begin{picture}(30,17)
    \triup
    \qbezier(0,0)(5,8.65)(10,17.3)
 \end{picture}
 \end{center}
 \end{minipage}
}

\newcommand{\QDMtupc}{\unitlength0.04em
 \begin{minipage}{35\unitlength}
 \begin{center}
 \begin{picture}(30,17)
    \triup
    \qbezier(20,0)(5,25.95)(10,17.3)
 \end{picture}
 \end{center}
 \end{minipage}
}

\newcommand{\QDMtdnb}{\unitlength0.04em
 \begin{minipage}{35\unitlength}
 \begin{center}
 \begin{picture}(30,17)
    \tridown
    \qbezier(20,0)(6,25.95)(10,17.3)
 \end{picture}
 \end{center}
 \end{minipage}
}

\newcommand{\QDMtdnc}{\unitlength0.04em
 \begin{minipage}{35\unitlength}
 \begin{center}
 \begin{picture}(30,17)
    \tridown
    \qbezier(20,0)(15,-8.65)(30,17.3)
 \end{picture}
 \end{center}
 \end{minipage}
}

\newcommand{\QDMtriup}{\unitlength0.04em
 \begin{minipage}{35\unitlength}
 \begin{center}
 \begin{picture}(30,17)
    \triup
    \put(0,0){\line(1,0){20}}
    \qbezier(0,0)(5,8.65)(10,17.3)
    \qbezier(20,0)(5,25.95)(10,17.3)
 \end{picture}
 \end{center}
 \end{minipage}
}

\begin{document}

\title{Single hole and vortex excitations in the doped Rokhsar-Kivelson quantum dimer model
on the triangular lattice}

\author{Hugo Ribeiro}

\author{Samuel Bieri}
\email[]{samuel.bieri@epfl.ch}

\author{Dmitri Ivanov}

\affiliation{Institute of Theoretical Physics, Ecole Polytechnique
F\'{e}d\'{e}rale de Lausanne (EPFL), CH-1015 Lausanne, Switzerland}


\date{August 3, 2007}

\begin{abstract}
We consider the doped Rokhsar-Kivelson quantum dimer model on the
triangular lattice with one mobile hole (monomer) at the
Rokhsar-Kivelson point. The motion of the hole is described by two
branches of excitations: the hole may either move with or without a
trapped Z$_2$ vortex (vison). We perform a study of the hole
dispersion in the limit where the hole hopping amplitude is much
smaller than the interdimer interaction. In this limit, the hole
without vison moves freely and has a tight-binding spectrum. On the
other hand, the hole with a trapped vison is strongly constrained
due to interference effects and can only move via higher-order
virtual processes.
\end{abstract}

\maketitle

\section{Introduction}

Quantum dimer models in two dimensions are an active field of
research due to their possible relevance to high-temperature
superconductivity\cite{RK87,RK88,kivelson89} and to frustrated
magnetism,\cite{Read-Sachdev} in the framework of the
resonating-valence-bond (RVB) construction.\cite{anderson87} It was,
however, realized that the RVB liquid state of the original
Rokhsar-Kivelson (RK) dimer model on the square lattice is
critical\cite{fisher6163} and does not provide a stable phase. A
search for a liquid with topological order\cite{wen91} continued
with quantum dimer models on nonbipartite lattices,\cite{ms01_2} in
particular on the triangular lattice. At a special value of the
coupling constants (the RK point), the RK dimer model on the
triangular lattice is proven to have exponentially decaying
correlations,\cite{ms01,Ioselevich} and a gapped excitation spectrum
is found numerically.\cite{dima04} Furthermore, there is now
convincing numerical evidence that this gapped liquid phase is
stable and extends within a finite parameter range.\cite{arnaud}

In the RVB scenario of high-temperature superconductivity, the
superconductivity emerges upon doping a liquid of singlet bonds with
charge carriers.\cite{anderson87,lee06,pvrvb,foot} Very early it was
therefore proposed to study quantum dimer models containing
monomers\cite{RK88,kivelson89} (see
Refs.~\onlinecite{syl04,poilblanc06,castelnovo07,arnaudPhaseSep,
fradkin07} for more recent studies). When analyzing a hole in the
background of dimer or spin liquids, one should take into account
vortexlike excitations inherent in topological
liquids.\cite{kivelson89,RC89} These topological excitations were
dubbed {\it visons} in the context of the corresponding Z$_2$ gauge
theory.\cite{SF} There is now evidence that visons constitute the
gapped excitations of the undoped triangular-lattice quantum dimer
liquid.\cite{dima04,arnaud} It is known that visons can bind to
holes and change their statistics.\cite{kivelson89,RC89}

%It was realized that a hole in the background of dimer or spin
%liquids may lead to vortex-like excitations, which change the
%statistics of the holes.\cite{kivelson89,RC89} These topological
%excitations were dubbed {\it visons} in the context of the
%corresponding Z$_2$ gauge theory.\cite{SF} There is now evidence
%that visons constitute the gapped excitations of the {\it undoped}
%triangular-lattice quantum dimer liquid.\cite{dima04,arnaud}

%In the construction of trial RVB wavefunctions for doped spin
%systems, such vortex-like excitations were argued to be
%energetically favored in some cases.\cite{RC89}

In this work, we study the properties of a single-hole excitation in
the case of a doped triangular-lattice quantum dimer model at the RK
point and illustrate the hole-vison binding. We find two branches of
excitations: one for the hole itself and the other for a hole-vison
bound state. The energy-momentum dispersion for both branches of
excitations is computed in the perturbative regime of small hole
hopping. The combined effects of lattice frustration and Z$_2$ flux
lead to a quadruply degenerate vison-hole branch and reduced
bandwidth. These results may have interesting implications for RVB
physics.

This paper is organized in the following way. In
Sec.~\ref{Section:model}, we introduce the model and show the
existence of two branches of excitations. In
Sec.~\ref{Section:non-vison}, we calculate perturbatively the
dispersion of the non-vison branch of the hole excitation. In
Sec.~\ref{Section:vison}, the dispersion of the vison branch is
calculated. Finally, in Sec.~\ref{Section:summary}, we summarize and
discuss our results.

\section{Doped dimer model, topological sectors, and two branches
of excitations}
\label{Section:model}

We consider the quantum dimer model on the triangular lattice doped
with mobile holes. We choose the simplest form of the hole-hopping
term which involves rearrangement of one dimer. The Hamiltonian
reads
\begin{eqnarray}
\label{Hfull} && H_{RK+hole}\nonumber = \\ &&\sum_{\rhombi}\left[-t \left(\ket{\QDMts}\bra{\QDMtz} + \mathrm{h.c.}\right) +v \left(\ket{\QDMtz}\bra{\QDMtz}+ \ket{\QDMts}\bra{\QDMts}\right)\right]\nonumber \\
&& + \sum_{\QDMtriup}\left[ -s \left( \ket{\QDMtupa}\bra{\QDMtupc} +
\mathrm{h.c.} \right) + 2 u \ket{\QDMtupb}\bra{\QDMtupb} \right] \nonumber \\
&& = H_{RK} + H_{s}\; ,
\end{eqnarray}
where the first sum is performed over all three orientations of
rhombi and the second sum is over both up and down triangles and
over all three possible positions of the hole on the triangle.

We consider the model [Eq.~\eqref{Hfull}] at the RK point, $t=v=1$,
in the sector with a single hole. At $s = u \geq 0$, the Hamiltonian
has the usual ``supersymmetric'' properties of the RK point: its
ground state is exactly known and given by the equal-amplitude
superposition of all possible states,\cite{RK88} and the quantum
mechanics can be mapped onto a classical stochastic dynamics in
imaginary time.\cite{henley04} We further consider the hole term
$H_s$ as a perturbation in $s\ll 1$, $u\ll 1$. To simplify the
formulae, we assume $u=s \geq 0$, but our results are trivially
extendable to $u\ne s$.

In the unperturbed Hamiltonian $H_{RK}$, the position of the hole
$x$ is a static parameter. We consider the hole on the infinite
lattice (or, equivalently, on a large finite lattice far from the
boundary). In such a setup there are two degenerate ground states of
$H_{RK}$ for each hole position. They correspond to two disconnected
(topological) sectors $\mathcal{H}^{\pm}(x)$ of the Hilbert space,
characterized by the values $\pm 1$ of the vison operator
\begin{equation}
V(x) = (-1)^\text{No. of dimers intersecting $\Gamma_x$},
\label{vison-definition}
\end{equation}
for some contour $\Gamma_x$ connecting the hole position $x$ to
infinity (in a finite system to the
boundary).~\cite{kivelson89,RC89} The corresponding ground states
are given by the sums over all dimer coverings in the respective
topological sector and are denoted as $\psi_0^{\pm}(x)$. Note that
while the labeling $\pm$ of the two sectors $\mathcal{H}^{\pm}(x)$
depends on the choice of the contour $\Gamma_x$, the sectors
themselves do not. Changing the contour $\Gamma_x$ amounts to
possible interchanges $\mathcal{H}^{+}(x) \leftrightarrow
\mathcal{H}^{-}(x)$ and, therefore, $\psi_0^{+}(x) \leftrightarrow
\psi_0^{-}(x)$. This ambiguity reflects the Z$_2$ degree of freedom
in labeling the topological sectors, and it will play an important
role in the motion of the hole with a trapped vison. Technically,
this Z$_2$ gauge may be fixed by specifying (arbitrarily), for each
$x$, a reference dimer covering which belongs to $\mathcal{H}^+(x)$.

The two topological sectors $\mathcal{H}^{\pm}(x)$ differ by the
parity of the dimer intersection at infinity and hence are
indistinguishable by any local operator (since all correlation
functions are short-ranged in the RK model on the triangular
lattice~\cite{ms01,Ioselevich}). Therefore, for excitations obtained
from the ground states by local operators (in the vicinity of $x$),
one can establish a one-to-one linear mapping between the states in
$\mathcal{H}^{+}(x)$ and in $\mathcal{H}^{-}(x)$. Taking odd and
even combinations of the corresponding states
$\psi^{+}(x)\pm\psi^{-}(x)$, we obtain the decomposition of the
Hilbert space into even and odd sectors $\mathcal{H}^{e,o}(x)$.
Those even and odd sectors correspond to the non-vison and vison
sectors of excitations, respectively, introduced in
Ref.~\onlinecite{dima04}.

The key observation for our discussion is that the Hamiltonian
[Eq.~\eqref{Hfull}] preserves the decomposition into
$\mathcal{H}^{e,o}(x)$ at every point $x$. While it is obviously
true for $H_{RK}$ and the potential part of $H_s$, one can also
easily check that the hopping part of $H_s$ does not have matrix
elements between $\mathcal{H}^e(x)$ and $\mathcal{H}^o(x')$ for
neighboring sites $x$ and $x'$. Hence, the excitations of the moving
hole can also be classified into two branches: the non-vison branch
[contained in $\oplus_x \mathcal{H}^e(x)$] and the vison branch
[contained in $\oplus_x \mathcal{H}^o(x)$]. This splitting into even
(non-vison) and odd (vison) branches is a generic feature of any
perturbative mixing of topological sectors in quantum dimer models.

\section{Non-vison branch}
\label{Section:non-vison}

The energy spectrum of the non-vison branch can be easily calculated
to first order in the perturbative expansion in $H_s$. We can fix
the phases of all RK ground states
$\psi^e_0(x)=\psi^+_0(x)+\psi^-_0(x)$ by taking the linear
combination of all dimer coverings with the amplitude one (up to
normalization). Then, the problem of the moving hole maps onto the
tight-binding model with the hopping amplitude
\begin{equation}
t_1=-\bra{\psi^e_0(x)}H_s\ket{\psi^e_0(x')}
\end{equation}
for nearest-neighbor $x$ and $x'$ (here and in the following we
always assume normalized states). This amplitude may be converted
into an expectation value in the RK model with a static hole,
\begin{equation}
t_1 = 2s \braket{\psi_0(x)}{\QDMtupa} \braket{\QDMtupa}{\psi_0(x)}
=2s\frac{N_3}{N_1}\, ,
\end{equation}
where $N_1$ and $N_3$ are the numbers of dimer coverings with one
site and one three-site triangle removed, respectively. The ratio
$N_3/N_1$ is well defined in the limit of the infinite system and
can be computed numerically with a suitable method. We have
calculated this coefficient with a Monte Carlo simulation similar to
that in Refs.~\onlinecite{Ioselevich} and \onlinecite{dima04} (using
clusters of toroidal geometry with up to 17$\times$17 sites), with
the result $N_3/N_1 = 0.229\pm 0.001$.

Taking into account the potential term in $H_s$ and performing the
Fourier transformation in $x$, the dispersion of the hole without a vison
takes the form
\begin{equation}
E_{\bf k} = - 2 t_1(\cos k_1 + \cos k_2 + \cos k_3 - 3)\, ,
\label{spectrum-non-vison}
\end{equation}
where $k_1$, $k_2$, and $k_3$ are the projections of the vector ${\bf k}$
on the three lattice directions (with $k_1+k_2+k_3=0$).

\begin{figure}
\includegraphics[width=.25\textwidth]{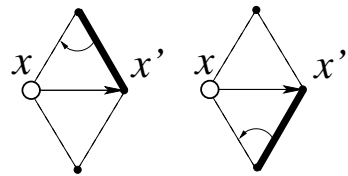}
\caption{ These two types of hopping processes set opposite
correspondences between the sectors $\mathcal{H}^\pm(x)$ and
$\mathcal{H}^\pm(x')$, and therefore cancel each other in
Eq.~\eqref{first-order-cancellation}.} \label{fig1}
\end{figure}

\section{Vison branch} \label{Section:vison}
The hopping of a hole with a trapped vison is more complicated. The
phases of the odd-sector ground states, $\psi^o_0(x) =
\psi_0^+(x)-\psi_0^-(x)$, cannot be synchronized invariantly for all
$x$, which reflects the frustration of the vison
motion.~\cite{dima04} The freedom of the Z$_2$ gauge [the contours
$\Gamma_x$ in Eq.~\eqref{vison-definition} or, equivalently, the
reference dimer configuration for each $x$] corresponds to the
choice of the overall sign for the states in ${\mathcal H}^o(x)$.

Regardless of the chosen Z$_2$ gauge, the hopping amplitude vanishes
to first order,
\begin{equation}
\bra{\psi^o_0(x)} H_s \ket{\psi^o_0(x')}=0
\label{first-order-cancellation}
\end{equation}
for nearest-neighbors $x$ and $x'$. This can be seen as the
cancellation of the two types of hopping processes from $x$ to $x'$,
corresponding to two possible dimer flips (Fig.~1). Each of those
dimer flips maps each of $\mathcal{H}^\pm(x)$ into one of
$\mathcal{H}^\pm(x')$. The change in topological sector depends on
the chosen gauge, but the correspondence between the two sectors
$\mathcal{H}^\pm(x)$ and the two sectors $\mathcal{H}^\pm(x')$ is
opposite for the two types of flips.~\cite{hugo_thesis} As a result,
the corresponding processes connecting two ground states
$\psi^o_0(x)$ and $\psi^o_0(x')$ exactly cancel each other.

A nontrivial hopping appears only to higher order in perturbation
theory for some trajectories. The second-order hopping amplitude
\begin{equation}
t_2=\sum_{x',n\neq 0} \frac{1}{E_n}
\bra{\psi^o_0(x)}H_s\ket{\psi^o_n(x')}
\bra{\psi^o_n(x')}H_s\ket{\psi^o_0(x'')} \label{t2-definition}
\end{equation}
involves excitations $\psi^o_n(x')$ of $H_{RK}$ with energies $E_n$.

\begin{figure}
\includegraphics[width=.35\textwidth]{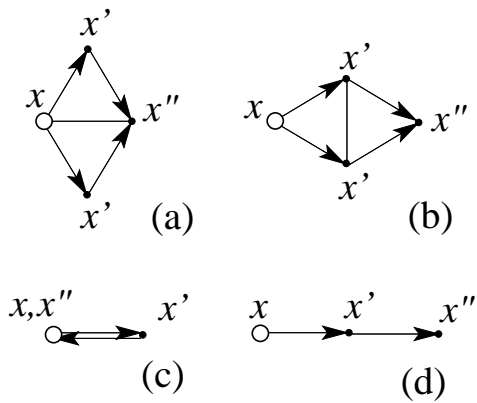}
\caption{The two trajectories of the hole exactly cancel each other
in the second order of the perturbation theory
[Eq.~\eqref{t2-definition}] for (a) nearest-neighbor and (b)
next-nearest-neighbor hoppings. The nonvanishing second-order terms
are (c) on-site and (d) next-next-nearest-neighbor hoppings.}
\label{fig2}
\end{figure}

Similarly to the cancellation of the nearest-neighbor hopping
amplitude to first order in perturbation, one can show the
cancellation to second order of the hopping processes $x\to x'\to
x''$ connecting nearest-neighbor and next-nearest-neighbor sites
[processes (a) and (b) in Fig.~\ref{fig2}]. One can verify that, in
those cases, processes symmetric with respect to the line $xx''$
exactly cancel each other.

The only nontrivial hopping in the second order occurs for
trajectories $x\to x'\to x''$ involving two links along the same
direction [i.e., for the on-site energy correction and for the
next-next-nearest-neighbor hopping, processes (c) and (d) in
Fig.~\ref{fig2}]. The corresponding next-next-nearest-neighbor
hopping amplitude [Fig.~\ref{fig2}(d)] to second order in
perturbation [Eq.~\eqref{t2-definition}] may be expressed via
dynamic correlation functions in the RK model with a static hole at
position $x'$,
\begin{eqnarray}
t_2&=&\int_0^\infty dt \bra{\psi^o_0(x)} H_s e^{-H_{RK} t} H_s
\ket{\psi^o_0(x'')}
\nonumber\\
&=&s^2\int_0^\infty dt\, I(t)\, , \label{t2-calculation}
\end{eqnarray}
where
\begin{equation}
I(t)=\bra{\psi_0(x')} P_{xx'} e^{-H_{RK} t} P_{x'x''}\ket{\psi_0(x')}
\end{equation}
and
\begin{eqnarray}
P_{xx'}&=& \ket{\QDMtupb}\bra{\QDMtupb} - \ket{\QDMtdnb}\bra{\QDMtdnb}\nonumber \\
P_{x'x''}&=& \ket{\QDMtdnc}\bra{\QDMtdnc}-\ket{\QDMtupc}\bra{\QDMtupc} \, .
\end{eqnarray}
The dynamic correlation function $I(t)$ is well defined in the limit
of infinite system size and does not depend on the topological
sector in this limit. It may be computed with a classical Monte
Carlo method as in Ref.~\onlinecite{dima04}. Using clusters of
toroidal geometry and up to 17$\times$17 sites, we find
$\int_0^\infty dt\; I(t) = -1.51\pm 0.08$ (observe that it is
negative).

\begin{figure}
\includegraphics[width=.2\textwidth]{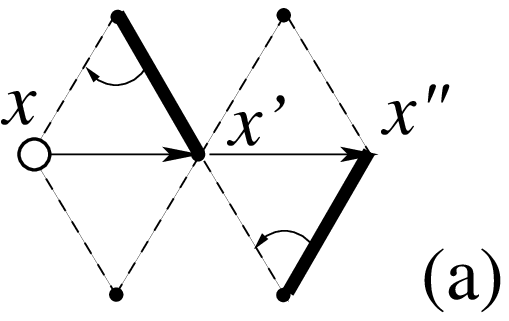}
\vskip 0.5cm
\includegraphics[width=.4\textwidth]{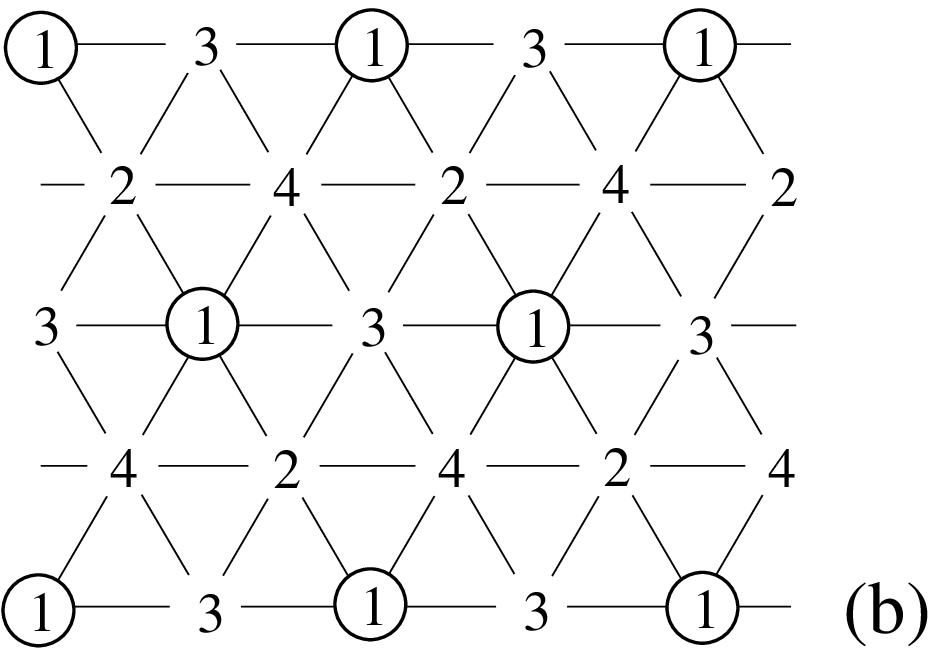}
\caption{ (a) We fix the relative gauge at the
next-next-nearest-neighbor sites by relating the reference
configurations in $\mathcal{H}^+(x)$ and $\mathcal{H}^+(x'')$ via
two consecutive dimer flips on opposite sides of the line $xx'x''$.
(b) The four sublattices connected by the hopping of the vison-hole
bound state. This composite excitation can only hop by multiples of
two lattice periods.} \label{fig3}
\end{figure}

Note that the sign of $t_2$ in Eq.~\eqref{t2-calculation}]
corresponds to a particular relative gauge choice at points $x$ and
$x''$: the reference dimer coverings at $x$ and $x''$ are connected
by two dimer flips on {\it opposite} sides of the line $xx'x''$
(Fig.~\ref{fig3}). One can show that this local gauge convention for
any two sites separated by two lattice periods can be consistently
extended to a global gauge on the sublattice of such sites (with the
period of this sublattice equals twice that of the original
lattice). There are four such sublattices (Fig.~\ref{fig3}), and the
hole-vison excitation hops on each of them independently, without a
possibility to cross over to another sublattice. The resulting
dispersion relation is that of the tight-binding model with the
doubled lattice constant and the hopping amplitude given by Eq.~
\eqref{t2-calculation},
\begin{equation}
E_{\bf k}^{\rm (v)}= -2t_2 (\cos 2k_1 + \cos 2k_2 + \cos 2k_3)
+ \varepsilon_0 \, .
\label{hole-vison-dispersion}
\end{equation}
The on-site energy $\varepsilon_0$ is equal to that in the non-vison
sector in Eq.~\eqref{spectrum-non-vison}. To leading order in $s$,
it is given by $\varepsilon_0=6 t_1$.

The hole-vison excitations with dispersion
\eqref{hole-vison-dispersion} are quadruply degenerate (by
sublattice) for each value of ${\bf k}$ in the Brillouin zone of the
doubled lattice. While we have explicitly demonstrated this
degeneracy to second order, it can be extended to all orders of
perturbation theory. In fact, this degeneracy is determined by the
symmetries of the original Hamiltonian [Eq.~\eqref{Hfull}] in the
vison sector and can thus be promoted from a perturbative argument
to the exact spectrum. The exact degeneracy can be proven using the
translational invariance of the Hamiltonian, together with the
symmetry under point inversion (rotation by $\pi$) and time
reversal. Physically, this degeneracy can be understood as the
cancellation of virtual processes for the flux-carrying excitation
on the frustrated triangular lattice.

Finally, let us note that, while our derivation of the vison-hole
spectrum was formally done at the RK-point, its form and degeneracy
are the same in the whole liquid phase away from the RK point
(estimated to extend to the region $0.8\lesssim\frac{v}{t}\leq 1$ in
Ref.~\onlinecite{arnaud}), provided the hole hopping is small. Only
the numerical coefficients in the hopping amplitudes $t_1$ and $t_2$
get modified in this case. Furthermore, our results equally apply
when more than one hole is present in the system as long as the
holes are sufficiently far apart and do not interact with each
other.

\section{Summary} \label{Section:summary}

In this paper, we have calculated the dispersion of a single mobile
hole in the RVB liquid phase of the doped RK quantum dimer model on
the triangular lattice. We find two branches of excitations: one for
the bare hole and the other for a hole-vison bound state. The
effective motion of the hole-vison state is strongly modified by the
Z$_2$ flux associated with the vison. Interference effects due to
lattice frustration reduce the bandwidth of this type of excitation
and lead to additional degeneracies. These are general properties,
which should be observed in any doped Z$_2$ RVB liquid on frustrated
lattices.

In our specific model [Eq.~\eqref{Hfull}], the energy of a {\it
static} ($s=0$) vison-hole bound state equals that of a hole without
a vison. In other words, the vison does not cost any energy if
placed in a hole (while in the bulk, its energy is a finite fraction
of $t$, see Ref.~\onlinecite{dima04}). In the limit of a small
hopping amplitude $s$, the energy of a static excitation is split,
with the bandwidth proportional to $s$ for the bare hole and to
$s^2/t$ for the hole-vison bound state. As a result, the two
branches intersect each other, with the minimum of energy (the
ground state) corresponding to the hole without a vison. For some
${\bf k}$ in a region close to the boundary of the Brillouin zone,
the vison-hole bound state is lower in energy than the bare hole. In
a more general quantum dimer model (or in other RVB-type systems),
however, one may imagine the situation where the hole-vison bound
state constitutes the ground state (in our dimer model, this may be
achieved, for example, by adding ring exchange of dimers around a
hole). In such a case, the doped holes spontaneously generate
visons, which, in turn, may lead to further interesting effects,
e.g., the modification of the statistics of
holes.\cite{RC89,kivelson89}

\acknowledgements We thank George Jackeli and Mike Zhitomirsky for
useful discussions. This work was supported by the Swiss National
Science Foundation.


\begin{thebibliography}{99}
\bibitem{foot}
While high-temperature superconductors are now believed to be
described by a long-range RVB liquid [Ref.~\onlinecite{pvrvb}],
promising short-range RVB states have been proposed recently for
several frustrated spin systems. See, e.g.,
Refs.~\onlinecite{ys06,mike05,francois06}.
\end{thebibliography}
\end{document}